\documentclass{PoS}

\usepackage{amssymb,amsmath}

\newcommand{\non}{\nonumber}
\newcommand{\half}{\tfrac{1}{2}}
\newcommand{\tr}{{\rm tr}}
\newcommand{\UA}{U$_{\rm A}$}
\newcommand{\m}{ \mathcal{M} }

\title{Chiral phase transition in a random matrix model with three flavors}

\ShortTitle{Chiral phase transition in a random matrix model with three flavors}

\author{Hirotsugu Fujii\\
Institute of Physics, University of Tokyo, Tokyo 153-8902, Japan}

\author{Munehisa Ohtani\\
Physics Department, School of Medicine, Kyorin University, Tokyo 181-8611, Japan
}

\author{\speaker{Takashi Sano}\\
Department of Physics, University of Tokyo, Tokyo 113-0033, Japan\\
Institute of Physics, University of Tokyo, Tokyo 153-8902, Japan\\
E-mail:\email{tsano@nt1.c.u-tokyo.ac.jp}}

\abstract{
The chiral phase transition in the conventional random matrix model 
is the second order in the chiral limit, irrespective of the
number of flavors $N_f$, because it lacks the 
U$_{\rm A}(1)$-breaking determinant interaction term.
Furthermore, it predicts an unphysical value of zero for the topological
susceptibility at finite temperatures.
We propose a new chiral random matrix model
which resolves these difficulties by
incorporating the determinant interaction term 
within the instanton gas picture. 
The model produces a  second-order transition
for $N_f=2$ and a first-order transition for $N_f=3$, and recovers
a physical temperature dependence of the topological susceptibility.
}

\FullConference{The XXVII International Symposium on Lattice Field Theory - LAT2009\\
		 July 26-31 2009\\
		 Peking University, Beijing, China}

\begin{document}

\section{Introduction}

One of the prominent features of non-perturbative QCD is spontaneous
breaking of chiral symmetry in the light quark sector.
According to the Banks-Casher relation \cite{BanksC1979},
the order parameter $\langle \bar{q} q \rangle$ is related
to the Dirac spectral density at zero eigenvalue
in the thermodynamic limit.
Within the so-called $\epsilon$ regime characterized by the system
size $L^4$ such that $m_\pi \ll 1/L \ll$ 1 GeV, the QCD partition
function is dominated by the constant pion field configurations.
In this regime, universal properties of the
Dirac eigenvalue distribution near zero are legitimately analyzed in a
chiral random matrix (ChRM) theory \cite{ChRM}, 
where the kinetic term of the Dirac operator is discarded and the 
complexity of the gauge field dynamics is represented by treating 
the Dirac operator as a random matrix of constant modes.
The matrix size of the constant modes $2N$ is considered to be
proportional to the system volume.  By taking a large volume limit
$N\to \infty$, away from 
the $\epsilon$ regime, one can study thermodynamics of the ChRM
theory as a schematic mean-field model for QCD.
One finds the ground state of the model in the chirally broken
phase and can explore the model phase diagram by introducing the
temperature $t$ \cite{Jackson:1995nf}
and the quark chemical potential $\mu$ \cite{Stephanov:1996ki}.

There are two drawbacks, however, in the ChRM model,
concerning the U$_{\rm A}$(1) symmetry.
Firstly, the U$_{\rm A}$(1)-breaking determinant term is missing
in the effective action of the model, which
consequently predicts a second-order phase
transition at finite temperature for any number of the
quark flavors $N_f$. 
Secondly, the topological susceptibility is suppressed unphysically 
to vanish at finite temperatures.

In an earlier work \cite{JanikNZ97}, the appropriate form of
the ChRM model with a U$_{\rm A}$(1)-breaking term is speculated
from a quark model with the determinant interaction in 0+1 dimensions.
In addition to the constant modes in the conventional ChRM model, 
new constant modes are introduced so as to reproduce the determinant
interaction.  
These new modes are considered to be associated with instantons
and called topological zero modes.
However the effective potential of the starting quark model is
unbounded from below for $N_f=3$, and therefore
no physical ground state exists.

In this paper we propose a new ChRM model \cite{Sano:2009wd}
changing the distribution of the topological
zero modes, which results in the determinant term appearing under a
logarithm. 
Then the effective potential becomes stable for any $N_f$,
and describes a second-order phase
transition for $N_f=2$ while a first-order transition for 
$N_f=3$. Moreover, the topological susceptibility shows physical
behavior as a function of temperature in our model.

\section{Chiral random matrix model at finite temperature}\label{s:conv}

In analogy with the QCD partition function, 
the ChRM model with $N_f$ flavors of mass $m_f$ is
defined as an average of the quark determinants
\cite{Jackson:1995nf}  
\begin{align}
Z_\nu = \int dW e^{-N\Sigma^2 \tr W^\dagger W} \prod_{f=1}^{N_f} \det(D+m_f) 
\; ,
\end{align}
where the Dirac operator has been replaced with an anti-Hermite 
matrix of constant modes
\begin{align}
D=\left ( \begin{array}{cc} 0 & iW+it{\mathbf{1}_{N-|\nu|/2}}
\\ iW^\dagger+it{\mathbf{1}_{N-|\nu|/2}} & 0 \end{array}
\right )
\end{align}
with an $(N + \nu/2) \times (N - \nu/2)$ complex matrix $W$.
The Gaussian random distribution of $W$ is a simple realization of 
the complexity of the gluon dynamics.
Chiral symmetry is retained
as a fact $\{D, \gamma_5 \}=0$ with 
$\gamma_5 ={\rm diag}(1_{N+\nu/2},-1_{N-\nu/2})$. 
The temperature effect has been introduced in the Dirac operator
with a constant $t$, which may be interpreted as the lowest
Matsubara frequency $\pi T$. One can easily show that this matrix $D$
has $|\nu|$ exact zero eigenvalues, which are
interpreted as the zero modes accompanied by the topological
charge $\nu$.  
The complete partition function is obtained after the sum over 
$\nu$ weighted by the quenched topological
susceptibility $\tau$ of the pure gluon dynamics:
\begin{align}
Z_\theta 
= \sum_{\nu=-2N}^{2N} 
e^{-\frac{\nu^2}{2(2N)\tau}}e^{i\nu \theta}
Z_\nu
\; ,
\label{eq:conv}
\end{align}
where the $\theta$ angle has been introduced.

After rewriting the determinant with the fermion variables and
doing the Gaussian integral of $W$ in $Z_\nu$, we find a four-fermion vertex
interaction, which can be unfolded by introducing an 
$N_f \times N_f$ auxiliary variable $S \sim q_L^\dagger q_R$.
We perform the fermion integral 
to obtain an expression for $Z_\nu$ in terms of  $S$,
\begin{align}
Z_\nu
=&\int dS e^{-N\Sigma^2\text{tr}(SS^\dagger)}
\det \left (
(S+\m)(S^\dagger+\m )+t^2
\right )^{N-\tfrac{|\nu|}{2}}
\times
\begin{cases}
\det(S+\m)^{\nu} & (\nu \ge 0)\\
\det(S^\dagger+\m^\dagger)^{-\nu} & (\nu<0) \\
\end{cases}
\; ,
\label{eq:conv0}
\end{align}
where $\m={\rm diag}(m_1, \dots, m_{N_f})$.

In the thermodynamic limit we set $\nu=0$ and 
evaluate Eq.~(\ref{eq:conv0}) with the
saddle point equation. 
For $S \propto {\bf 1}_{N_f}$ and $\m=0$, the $N_f$ dependence is
factored out in the saddle point equation, implying a second order
phase transition for any number of $N_f$.
Furthermore, $Z_\nu$ is non-analytic in
$\nu$ as the integrand contains a term with $|\nu|$, which causes the
unphysical suppression of the topological susceptibility at finite
temperatures \cite{OhtaniLWH08}. Note that $|\nu|$ disappears when $t=0$.

\section{Chiral random matrix model with determinant interaction}

{\it Near-zero modes and topological zero modes --}
Let us first recall the instanton gas picture.
An isolated instanton is a localized object accompanying a
right-handed exact fermion zero mode.
In a dilute system of $N_+$ instantons and $N_-$ anti-instantons,
we expect $N_+$ right-handed and $N_-$ left-handed zero modes
even at finite temperatures.
In an effective theory at long distances, effects of the instantons
should be integrated out, which will result in \UA(1)-breaking effective
interactions.
The fundamental assumption in our model is the classification
of the constant modes into
the near-zero modes and the topological zero modes
\cite{JanikNZ97,Sano:2009wd}.
We deal with the $2N$ near-zero modes appearing in the conventional
models and additionally the $N_+ + N_-$
topological zero modes which we regard as the modes accompanied by
the instantons.
In our model, $N_\pm$ fluctuate according to the instanton distribution,
and the number of the exact zero modes is given by $\nu=N_+ - N_-$.
Eventually we sum over $N_+$ and $N_-$ assuming a certain distribution
with the mean value of ${\mathcal O}(N)$.

We write down a Gaussian ChRM model 
for definite numbers of zero modes as \cite{JanikNZ97,Sano:2009wd}
\begin{align}
Z^N_{N_+,\, N_-}=&
\int dA dB dX dY 
e^{-N \Sigma^2 {\rm tr}(AA^\dagger +BB^\dagger +XX^\dagger +YY^\dagger)}
\prod_{f=1}^{N_f} \det(D+m_f) 
\label{eq:fixNpf}
\end{align}
with
\begin{align}
D=
\left(
\begin{array}{cccc}
  0             &iA+it{\mathbf{1}_N} & 0     & iX  \\
  iA^\dagger+it{\mathbf{1}_N} & 0      & iY    & 0   \\
  0             & iY^\dagger & 0 & iB  \\
  iX^\dagger     & 0      & iB^\dagger & 0
\end{array}
\right)
\; ,
\end{align}
where the matrix $A$ corresponds to the near-zero modes of the
conventional model and the $N_+ \times N_-$ matrix $B$
represents the topological zero modes. The matrices
$X$ and $Y$ induce mixing among these modes. Note that 
the temperature term $t$ is introduced only for the near-zero modes.
Following the same steps as in Sec.~\ref{s:conv}, we find the sigma model
representation for this ChRM model, which is analytic in $N_\pm$ as well
as $N$ unlike Eq.~(\ref{eq:conv0}): 
\begin{align}
Z^N_{N_+,\, N_-}=&\int dS \; e^{-N\Sigma^2 \text{tr} S^\dagger S}
\; \det  \left [ (S+\m)(S^\dagger+\m^\dagger) +t^2 \right ]^N
\; \det  (S+\m)^{N_+} \; \det  (S^\dagger + \m^\dagger)^{N_-} 
 .
\label{eq:fixedia}
\end{align}

{\it Distribution of the topological zero modes --}
The complete partition function is obtained after
summing $Z_{N_+,\, N_-}^N$ over the instanton numbers $N_+$ and $N_-$.
Here we simply assume independent distributions $P(N_\pm)$ for
$N_+$ and $N_-$,
{\it i.e.},
\begin{align}
Z_\theta =&
 \sum_{N_+,\, N_-}
e^{i \nu \theta} P(N_+)P(N_-)Z_{N_+,\, N_-}^N 
=\int dS e^{-2N \Omega(S;~t, m, \theta)}
\; .
\label{eq:ia_sum}
\end{align}
Let us first consider $P(N_\pm)$ in a dilute instanton gas picture.
For a one-instanton configuration,
one may assign a weight $\kappa$ compared with a no-instanton
configuration, and multiply a factor $N \propto V$ taking into account
the integration over the instanton location.
For a configuration with $N_{+(-)}$ (anti-)instantons,
we then have a Poisson distribution
\begin{align}
P_{\rm Po}(N_{\pm})=\frac{1}{N_{\pm} !} (\kappa N)^{N_\pm}
\end{align}
where the factorial $N_{+ (-)} !$ appears as the symmetry factor.
The summation with $P_{\rm Po}(N_\pm)$ in Eq.~(\ref{eq:ia_sum}) results in
the exponentiation of the determinant term \cite{JanikNZ97}:
\begin{align}
\Omega_{\text{Po}}=&
\half \Sigma^2 {\rm tr}SS^\dagger 
-\half\ln \det \left [ (S+\m)(S^\dagger+\m^\dagger) +t^2 \right ]
-\half \kappa 
[e^{i\theta}  \det (S+\m) + e^{-i\theta}\det (S^\dagger +\m^\dagger)]
.
\label{eq:njlpf}
\end{align}
This determinant term is commonly incorporated in effective models as
the \UA(1) anomaly term.
However this potential is unbound for $N_f=3$ in the ChRM model
because the term $\det(S+\m) \sim \phi^3$ for large 
$S=\phi {\bf 1}_{N_f}$  dominates over the other terms
in $\Omega_{\text{Po}}$.

It should be noticed here that the fermion coupling distorts the
$N_\pm$ distribution itself.
With including the determinant term of the topological zero modes in
Eq.~(\ref{eq:fixedia}), the effective distribution for
$N_+$ reads
\begin{align}
\widetilde P_{\rm Po}(N_+)=\frac{1}{N_+ !} 
\left (\kappa N d\right )^{N_+}
\end{align}
with $d=|\det(S+\m)|$, and similarly for $N_-$.
This means that the average value of $N_\pm$ increases
indefinitely with increasing $d\sim \phi^{N_f}$ as
$\left < N_\pm \right > = \kappa N d$.
However,
the possibility of infinitely many degrees of freedom $N_\pm$
within a finite volume is inadequate 
for a regularized low-energy effective theory.
We need a cutoff for $N_\pm$.

Here we set explicitly a maximum
value of ${\cal O}(N)$ for $N_\pm$.
We split the finite space-time volume into $\gamma N$ cells with
$\gamma$ being a constant of ${\cal O}(1)$, and assign a probability
$p$ for a cell to be occupied by a single (anti-)instanton
and $(1-p)$ for a cell unoccupied.
This assumption results in the binomial distributions for $N_\pm$:
\begin{align}
P(N_\pm)=
\left (\begin{array}{c}
\gamma N \\ N_{\pm}
\end{array}
\right ) 
\; p^{N_\pm} (1-p)^{\gamma N-N_\pm}
.
\label{eq:bino}
\end{align}
For a small $p$ and a large $\gamma N$, the binomial distribution
$P(N_\pm)$ is accurately approximated with the Poisson
distribution with the mean $\gamma N p$. But it cannot be a good approximation
for a large $p$.
The binomial distribution provides a stringent upper bound $\gamma N$ for
the number of modes $N_\pm$,
in contrast to the Poisson distribution.
The corresponding effective potential for $S$ is found to be
\begin{align}
\Omega(S;t,m,\theta)=&
\half \Sigma^2 {\rm tr}SS^\dagger 
-\half\ln \det \left [ (S+\m)(S^\dagger+\m^\dagger) +t^2 \right ]
\non \\
&-\half \gamma 
\left [\ln \left(e^{i\theta}  \alpha \det (S+\m)+1 \right)
     + \ln \left(e^{-i\theta}  \alpha \det (S^\dagger +\m^\dagger)+1\right)
\right ]
\label{eq:thepf}
\end{align}
with $\alpha=p/(1-p)$.

\begin{figure}[tb]
\begin{center}
\includegraphics[width=0.3\textwidth,angle=270]{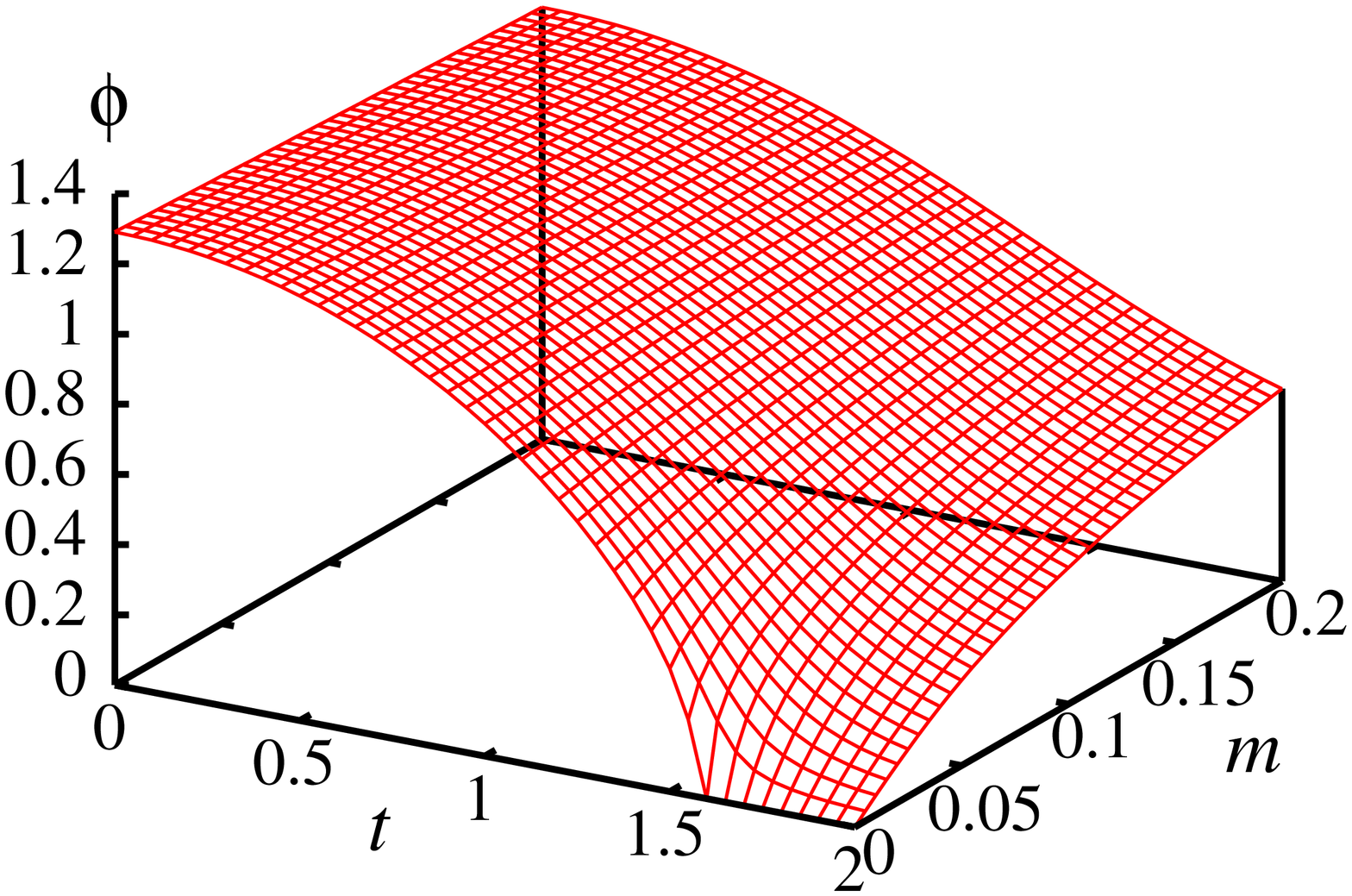}
\hfil
\includegraphics[width=0.3\textwidth,angle=270]{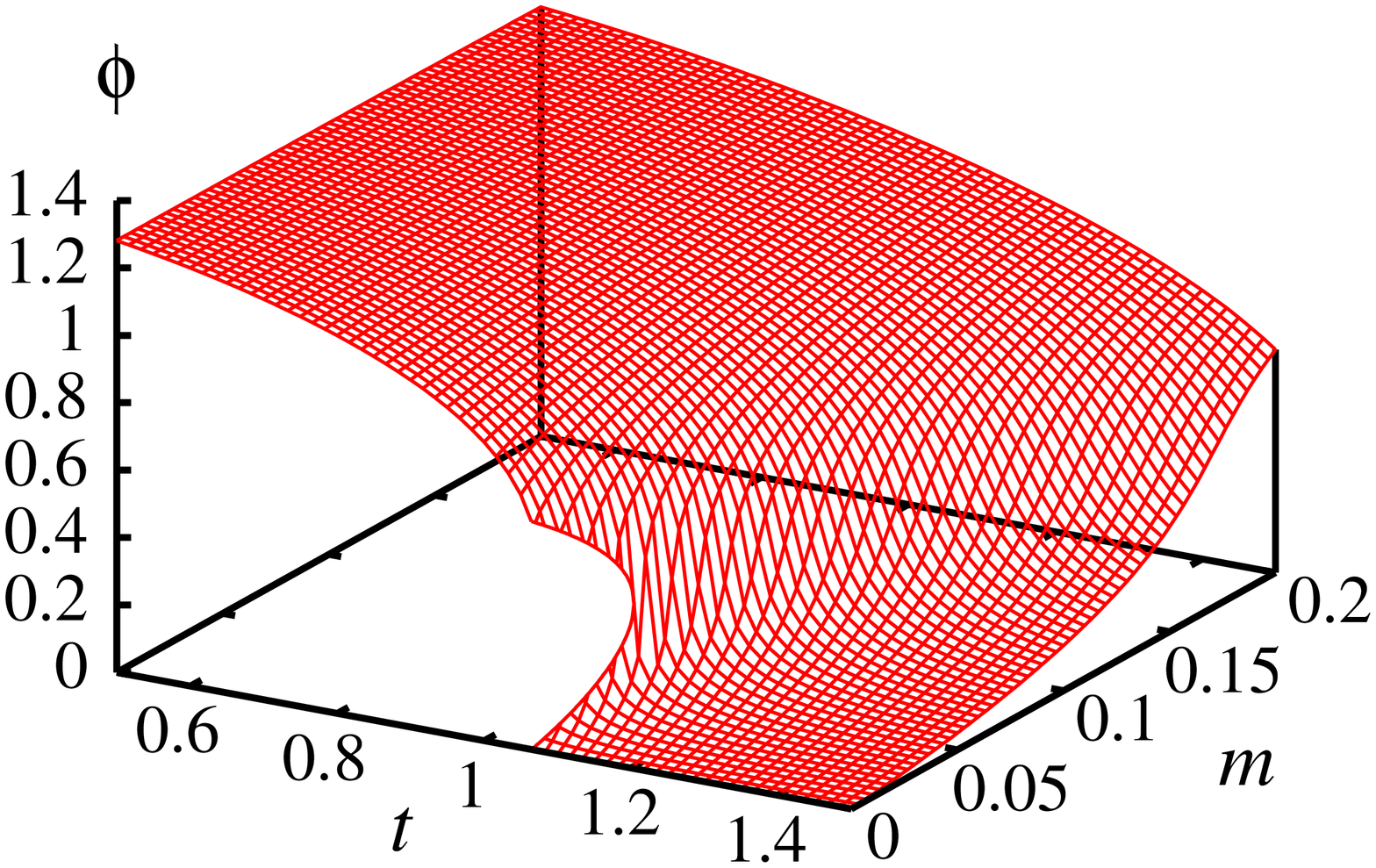}
 \caption{Chiral condensate $\phi$
as a function of $t$ and $m$ for $N_f=2$ (left) and 3 (right).
\label{fig:cond}}
\end{center}
\end{figure}

The variance of the topological charge $\nu = N_+ - N_-$
for the binomial distribution is computed
as $2N \tau = 2N \gamma p (1-p)$, where $\tau$ is the quenched
topological susceptibility. In the presence of the fermion coupling,
this susceptibility will be replaced with
\begin{align}
\tilde \tau=\gamma \tilde p (1-\tilde p)
=
\gamma \frac{\alpha d}
            {(\alpha d+1 )^2}
\; .
\label{eq:tildetau}
\end{align}

\section{Ground state and fluctuations}\label{s:num}

In this section we shall study ground state properties
of the system with equal mass,
$\m=m \mathbf{1}_{N_f}$, for simplicity.
Setting $S = \phi {\bf 1}_{N_f}$
with real $\phi$ and with $\theta=0$,
we obtain a simple form of the grand potential:
\begin{align}\label{eq:gtpot}
\Omega (\phi; t,m)=&
 \half N_f\Sigma^2 \phi^2
- \half N_f \ln \left[ (\phi + m)^2 + t^2 \right]
- \half \gamma \ln \left | \alpha (\phi + m)^{N_f} +1 \right |^2
\; .
\end{align}
The factor $N_f$ cannot be factored out in the potential $\Omega$
because of the anomaly term here.

\begin{figure}[tb]
\begin{center}
\includegraphics[width=0.4\textwidth]{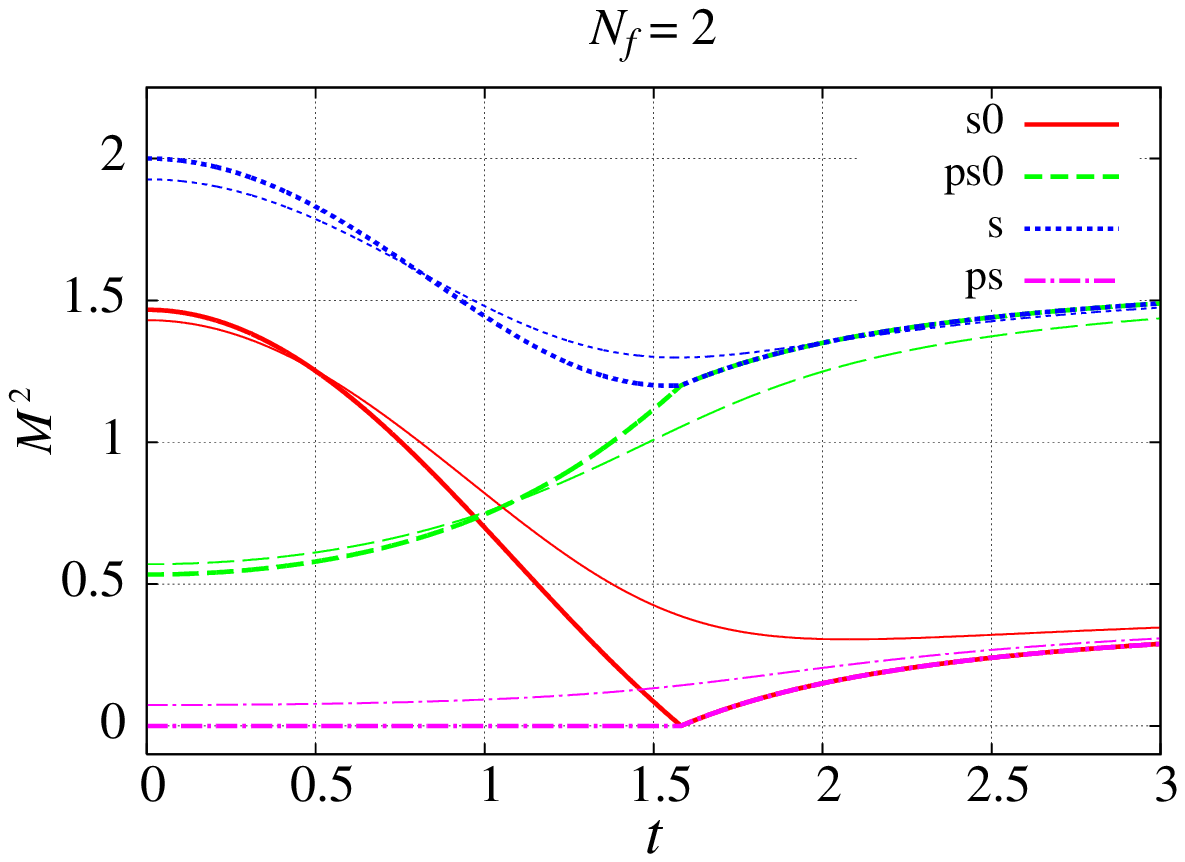}
\hfil
\includegraphics[width=0.4\textwidth]{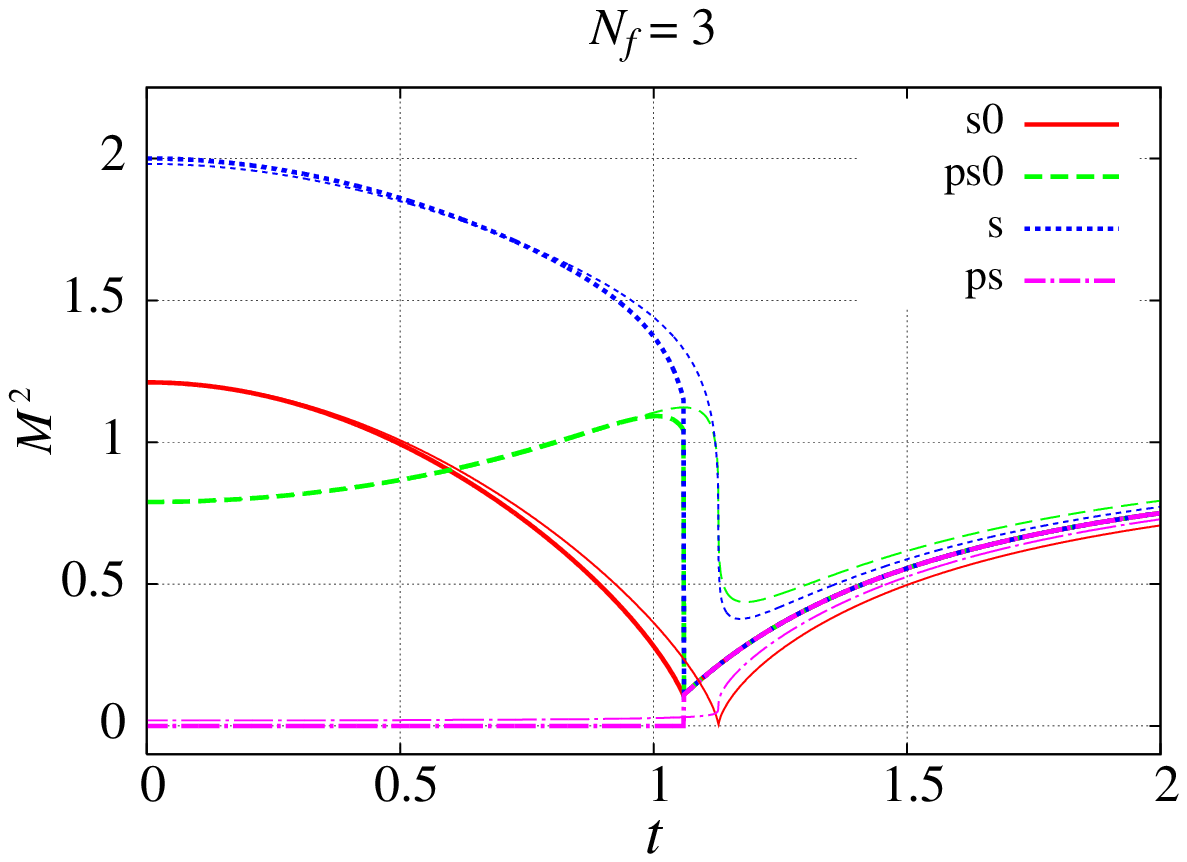}
\caption{Temperature dependence of
the mesonic masses in the flavor-singlet scalar (s0)
and pseudo-scalar (ps0) channels and in the flavor-nonsinglet
scalar (s) and pseudo-scalar (ps) channels
for $m=0$ (thick lines) and $m \ne 0$ (thin lines).
We set $m=0.1$ ($m_c=0.0265$) as nonzero quark mass for $N_f=2$ (3).
}
\label{fig:mass}
\end{center}
\end{figure}

In the thermodynamic limit we calculate the quark condensate
$\langle \bar q q \rangle \propto \phi$ for $\Sigma=1$, 
$\alpha=0.3$ and $\gamma=2$ in Fig.~\ref{fig:cond}.
The chiral phase transition is the second order for $N_f=2$ 
in the chiral limit, while it is the first order for $N_f=3$ 
and for $m<m_c=0.0265$.
The mesonic masses can be defined as 
$\Omega(S) = \Omega_0  + \tfrac{1}{2}M_{{\rm s}a}^2 \sigma^{a2}
+ \frac{1}{2}M_{{\rm ps}a}^2 \pi^{a2} + \cdots$
with $S=\phi+\lambda^a (\sigma^a+i\pi^a)/\sqrt{2}$  
parametrized with U($N$) generators
(tr($\lambda^a\lambda^b)=2\delta^{ab}$).
The temperature dependence of the masses are shown in Fig.~\ref{fig:mass}.
For pseudo-scalar flavor-nonsinglet masses we find the 
Gell-Mann--Oakes--Renner relation
\begin{align}
(\phi+m)^2 M_{\rm ps}^2 =m\Sigma^2 (\phi+m) \sim
-m \langle \bar q q \rangle
,
\end{align}
if we identify $\phi+m$ as the pion decay constant $f_\pi$.
On the other hand, the flavor singlet-masses
have an additional
contribution from the anomaly term as
\begin{align}
M_{\rm s0}^2 =&
M_{\rm s}^2  -\Delta M_0^2
\; ,
\qquad
M_{\rm ps0}^2 =
M_{\rm ps}^2 +\Delta M_0^2
\label{eq:ps0mass}
\end{align}
with 
$\Delta M_0^2 \equiv N_f \tilde \tau /(\phi+m)^2$.
The would-be Nambu-Goldstone mode becomes massive due to the
coupling to the \UA(1) interaction $\Delta M_0^2$,
and the mass gap is related to the (replaced) quenched susceptibility
$\tilde \tau$,
similarly to the Witten-Veneziano formula.

The topological susceptibility is obtained as
$\chi_{\rm top} =\partial^2\Omega(S(\theta);\theta)/\partial
\theta^2|_{\theta =0}$ with
the saddle point solution $S(\theta)=\phi+i\eta_0(\theta)\lambda^0/\sqrt{2}$
for small $\theta$, and we find \cite{OhtaniLWH08,Sano:2009wd}
\begin{align}
\frac{1}{ \chi_{\text{top}} }
=
\frac{1}{\tilde \tau } + \frac{1}{\tau_m}
\; .
\label{eq:chitop}
\end{align}
Here $\tilde \tau$ is the modified susceptibility defined in
Eq.~(\ref{eq:tildetau}).
The fermion coupling
screens $\chi_{\rm top}$ via the contribution
\begin{align}
\tau_m= \frac{\Sigma^2 m (\phi+m)}{N_f}
=\frac{M_{\rm ps}^2 (\phi+m)^2}{N_f}
\; .
\end{align}
Noting that the $\theta$ dependence can be absorbed into
the quark mass term as $m e^{i\theta/N_f}$, one can show 
the axial Ward identity
\begin{align}
\chi_{\text{top}} 
&=
-\frac{m^2}{N_f}\chi_{\text{ps}0}
-\frac{m}{N_f}\langle \bar q q \rangle
\; .
\end{align}
Since the pseudo-scalar meson in the flavor singlet channel
has nonzero mass (\ref{eq:ps0mass})
because of the \UA(1)-breaking term,
the pseudo-scalar singlet susceptibility remains finite
in the broken phase in the chiral limit. Thus for the small but nonzero
quark mass $m$, the decrease of $\chi_{\rm top}$ follows 
the chiral condensate $\langle \bar q q  \rangle \sim \phi$ with
increasing $t$, which is clearly observed in Fig.~3.

\begin{figure}[tb]
\begin{center}
\includegraphics[width=0.4\textwidth]{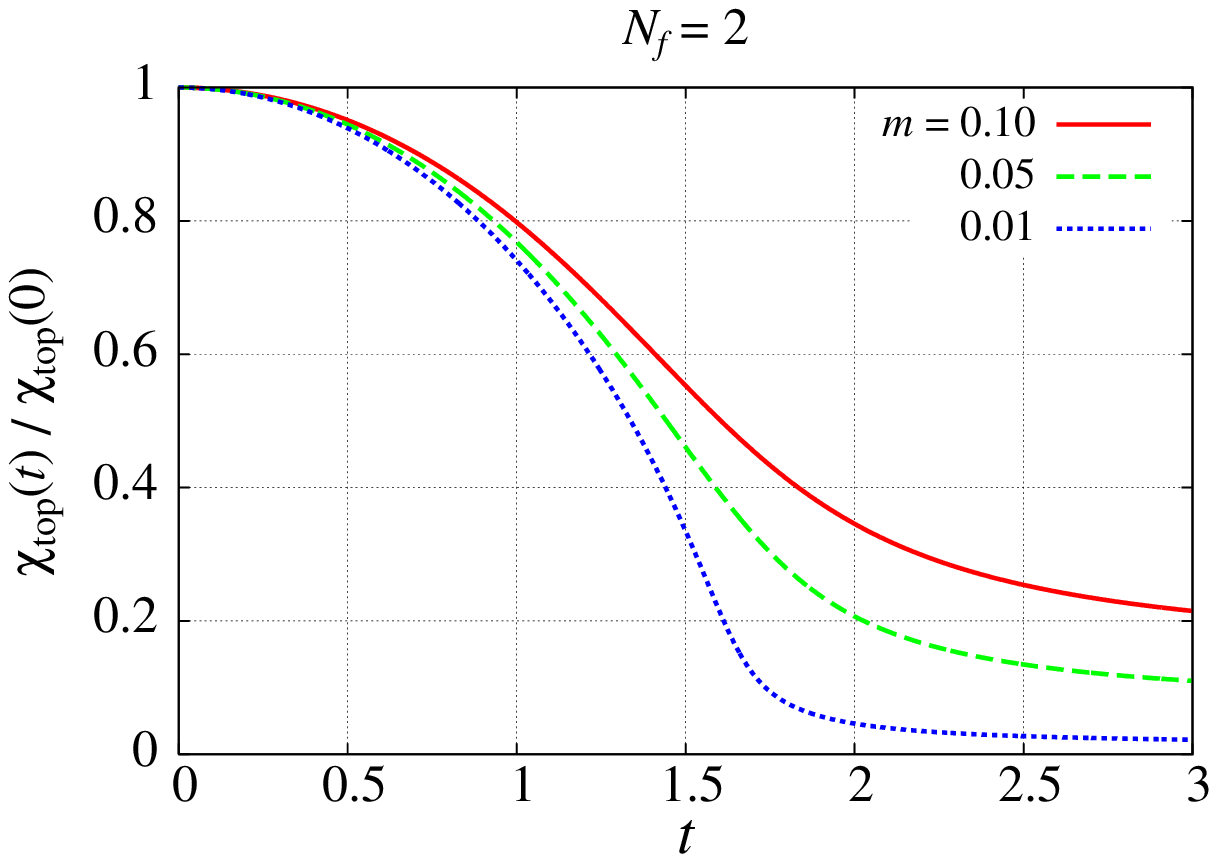}
\hfil
\includegraphics[width=0.4\textwidth]{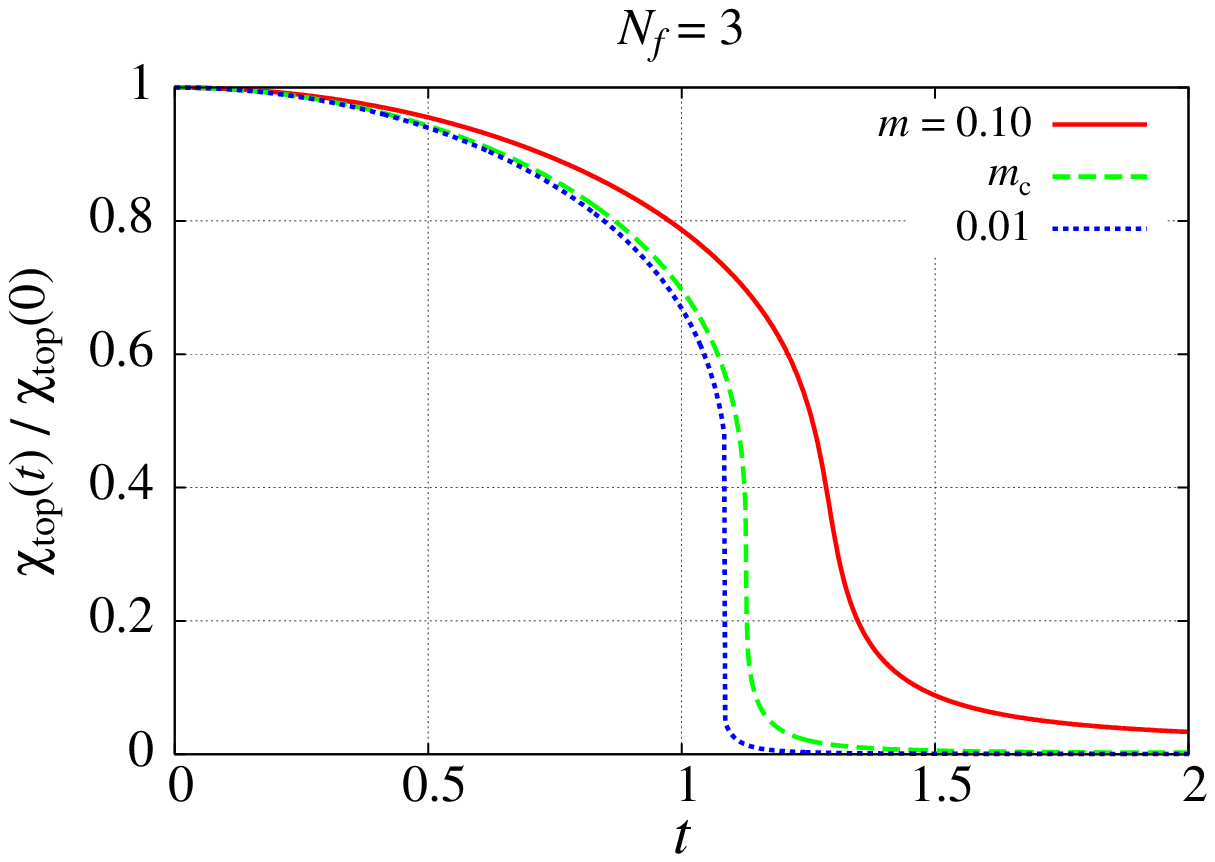}
 \caption{Temperature dependence of
topological susceptibility $\chi_{\rm top}(t)$ for $N_f=2$ and 3.
\label{fig:top}
}
\end{center}
\end{figure}

\section{Summary}\label{s:summary}

We have presented  a chiral random matrix model where
the determinant interaction is incorporated by summing up the topological
zero modes with the binomial distribution \cite{Sano:2009wd}.
The model has a stable ground state unlike the model
of \cite{JanikNZ97}, and describes the $N_f$ dependence of
the chiral phase transition. 
At the same time it reproduces the physical temperature dependence
of the topological susceptibility. 
This model can be applied to the 2+1 flavor case at finite temperature
and density~\cite{FujiiS}.

\vspace{2mm}
This work is supported in part by Grants-in-Aid 
of MEXT, Japan
(No.\ 19540269 and 19540273).

\end{document}